\documentclass[aps,amsmath,prb,a4paper,floatfix,twocolumn]{revtex4}
\usepackage{graphicx}
\usepackage{subfigure}
\usepackage{stackengine}
\usepackage{amsmath}
\usepackage{amsfonts}
\usepackage{amssymb}
\newcommand{\beq}{\begin{equation}}
\newcommand{\eneq}{\end{equation}}
\newcommand{\bea}{\begin{eqnarray}}
\newcommand{\enea}{\end{eqnarray}}
\newcommand{\met}{\frac{1}{2}} 
\newcommand{\freccia}{ \quad  \quad \mathbf{\Rightarrow} \quad }

\begin{document}
\title{Use of a spoof plasmon to optimize the coupling of  infrared radiation to Josephson Junction fluxon oscillations}
\author{ A.Tagliacozzo$^{1,2}$}
\author{S.De\! \!Nicola$^{2}$}
\author{D.Montemurro$^{1,2}$} 
\email{domenico.montemurro@unina.it}
\author{G.Campagnano$^{1}$}
\author{C.Petrarca$^{3}$} 
\author{C.Forestiere$^{3}$} 
\author{G.Rubinacci$^{3}$} 
\author{F.Tafuri$^{1,2}$} 
\author{G.P.Pepe$^{1,2}$}
\affiliation{$^{1}$  Dipartimento di Fisica, Universit\`{a}  degli Studi di Napoli Federico II, Via Cintia, I-80126 Napoli, Italy}
\affiliation{$^{2}$ CNR-SPIN, Monte S.Angelo via Cintia, I-80126 Napoli, Italy}
\affiliation{$^{3}$Department of Electrical Engineering, Universit\`{a}  degli Studi di Napoli Federico II, Via Claudio,I-80125,Napoli,Italy  }

\begin{abstract}
We show that Infrared radiation impinging onto a 1-d array of grooves drilled in the superconductor electrode of a long overlap junction can improve matching between  fluxon oscillations at $THz$ frequencies and  a spoof plasmon of  comparable wavelength. This  example proves that  metamaterials can be very helpful in  integrating superconductive and subwavelength optical circuits with optimized matching bridging the gap between infrared and microwave radiation.
\vspace{0.5cm}

\end{abstract}

\maketitle

\section{Introduction}
 Integrating superconductive and optical circuits in the infrared - microwave frequency range  would  boost  solid state design of quantum information processing  in a tremendous way\cite{heshami,held,giazotto}. By   engineering the optical absorption, Single Flux Quanta in long Josephson junctions can be manipulated\cite{mcDermott}. Connecting optical fibers or optical quantum memories\cite{lvovsky}, with  superconducting circuits\cite{stella} would increase performances and  operating speed as well as reduce power losses\cite{tlarkin}.
While integrated optics devices usually operate at the single-photon level\cite{politi,jaspan,heeres}, detection of surface plasmon or Surface Plasmon-Polariton (SPP) resonances induced by an evanescent field from a waveguide into a metal film appears as a different  promising method to keep  the  power  delivered  at the  interaction with the solid state device low and controlled\cite{sarid}.  Off-resonance, the evanescent non propagating field penetrating into the metal film is reflected back to the photodetector with minimal loss. At resonance, instead, energy is transferred to the metal film generating the surface plasmon mode that can be used to control a superconducting device. At present, detection of surface plasmon or SPP resonances is mostly being developed for biosensors and Surface Enhanced Raman Scattering substrates at visible and near infrared wavelength \cite{hunta}. Plasmonic photon sorters can be used for spectral and polarimetric imaging\cite{laux}. Surface-plasmons are already successfully used at very long wavelengths ( $60\div 160 \:\mu m $ wavelength) as a guiding solution for THz quantum cascade (QC) lasers\cite{williams}. Dielectric-based integrated optics is always limited in scaling by diffraction. Instead, optical generation of plasmon excitations uniquely offers a larger degree of confinement and therefore allows for the creation of structures smaller than the diffraction limit\cite{barnes,zhiwen}. SPP propagate in metamaterials (MM) obtained by etching metal surfaces with periodic subwavelength grooves or holes, at infrared frequencies\cite{pendry,garcia-vidal,maier}. Highly localized plasmon fields can be generated using ordered arrays of nanoparticles or nanohole arrays, instead of thin metal films\cite{forestiere}.   Changes in the environmental dielectric, will change the plasmon mode and shift the resonance to lower frequencies\cite{pendry1,linfang}. 
 \begin{figure}
\includegraphics[width=0.85\columnwidth]{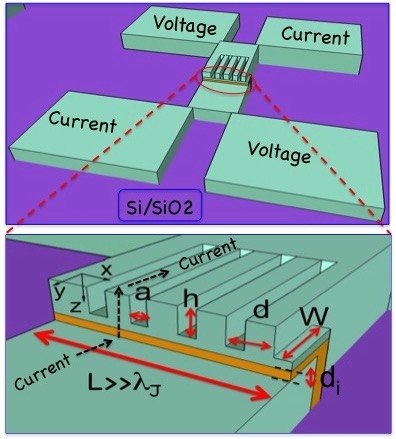}

\caption{ 
Layout of the proposed device based on a tunnel
Josephson junctions. The zooming of the junction shows
that the top contact is drilled with a regular array of
grooves (in light turquoise) on top of a superconducting sheet (dark turquoise), which is deposited onto the barrier (yellow sheet). The parameters involved are: 'd', the period of the 1-D groove lattice, '$d_i$', the insulating layer thickness (yellow part), 'a', 'h' , the distance between the pillars and their height, respectively. W, L and $\lambda_J$ represent width, length of the whole array of junctions and Josephson length, respectively.} 
\label{lioschema}
\end{figure}
 
 A large majority of existing MM designs rely on the use of metallic structures sitting on a dielectric substrate\cite{yeh,zhou}. However, as the frequency of operation is pushed higher toward the terahertz (THz), infrared, and visible, the Ohmic losses quickly render the current MM approaches impractical. Thus, a top priority is to reduce the absorption losses to levels suitable for device applications. This would require MM designs that do not depend solely on metallic structures and low temperature environment to prevent strong inherent vibrational absorption bands and the high skin-depth losses of the conductors\cite{yeh,zhou}. One approach would be to use  low power Josephson devices as the MM constituent media which allow dissipationless flow of electrical current\cite{gu,jung,butz,oh}.  Metamaterials with  rf SQUID meta-atoms have already been implemented to provide electromagnetically induced transparency (EIT)\cite{trepanier,lazarides,dzhang}.
 
 However, while plasmons belong to the high frequency optical band, Josephson junctions are usually controlled by shaped    free space microwaves   tones  ($ \lambda \sim 3\: cm$) at a  frequency: $\nu =  10^{11} Hz$, appropriate for fluxon  oscillations in long Josephson Junctions, which occur at a velocity which is about  $1/20 $th of that of light\cite{salerno}. This disparity in wave velocities makes it difficult to couple electromagnetic energy in and out of the junction region \cite{tinkham}.   
 
In this paper we propose to exploit subwavelength  optics to integrate infrared radiation with  fluxon  oscillations in  a long Josephson Junction\cite{lomdahl}. One of the  superconducting electrodes of a long Josephson Junction,  can be  modulated in shape, thus inducing periodic variation of the local critical current density which, in turn, is the source of radiative losses in the fluxon  dynamics. Infrared radiation impinging on the MM  electrode can generate  a SPP and appropriate choice of the MM geometry can trigger resonance between fluxon radiation in the insulating junction barrier and  the spoof plasmon in the infrared band\cite{golubov,filatrella}.  
 Such a trick would bridge the gap between infrared and microwave radiation in controlled Josephson systems.

 In Section II we briefly review  how infrared radiation can generate a SPP in the THz range by irradiating  a 1-d  subwavelength structure formed by an array of grooves drilled on the top of a  normal conductor electrode( see Fig.(\ref{lioschema}) for a sketch of the structure). We argue that there are limited consequences of the fact that the MM is fabricated in the superconducting electrode of the JJ.  In Section III we discuss how a fluxon generated in a long  overlap  Josephson Junction radiates in the junction as a consequence of the periodic modulation of its critical current density.  We show that it is possible to design the structure and the active circuit element in such a way that the energy dispersions of the fluxon and of the plasmon cross in the THz range. In Section IV we provide a simple model for the interaction between the radiation mode of the fluxon and the SPP mode. The interaction produces an  anticrossing of the two mode dispersions  and resonant  mixture of  the two modes  provides strong absorption.  In Section V the motion equation for the fluxon $\varphi (x,t) $  is extended by including the effects due to the  presence of the MM modulation and of the SPP interaction.  The latter generates a dissipative term which can be recognized as the third order derivative $\varphi _{xxt}$, dissipative '$\beta-$term'. Additionally, a forcing term arises, which strongly influences the fluxon dynamics, by increasing or decreasing its momentum, according to the phase of the applied perturbation. A simulation of the fluxon dynamics is reported and discussed in Section VI in the absence of dissipation. The pendulum motion of the fluxon between the junction edges can be highly  perturbed, and the fluxon can be  backscattered by a perturbation pulse. Increasing the forcing perturbation, multiple scattered waves are produced which interfere and produce beatings depending on the initial velocity of the fluxon.  However the shape of the principal kink is rather robust with the increase of the perturbation up to some critical velocity. Section VII collects the conclusions. Appendices A,B and C report some details on the derivation of the dissipative and forcing  terms.

\section{spoof surface plasmon dispersion}

\begin{figure}
\includegraphics[height=40mm]{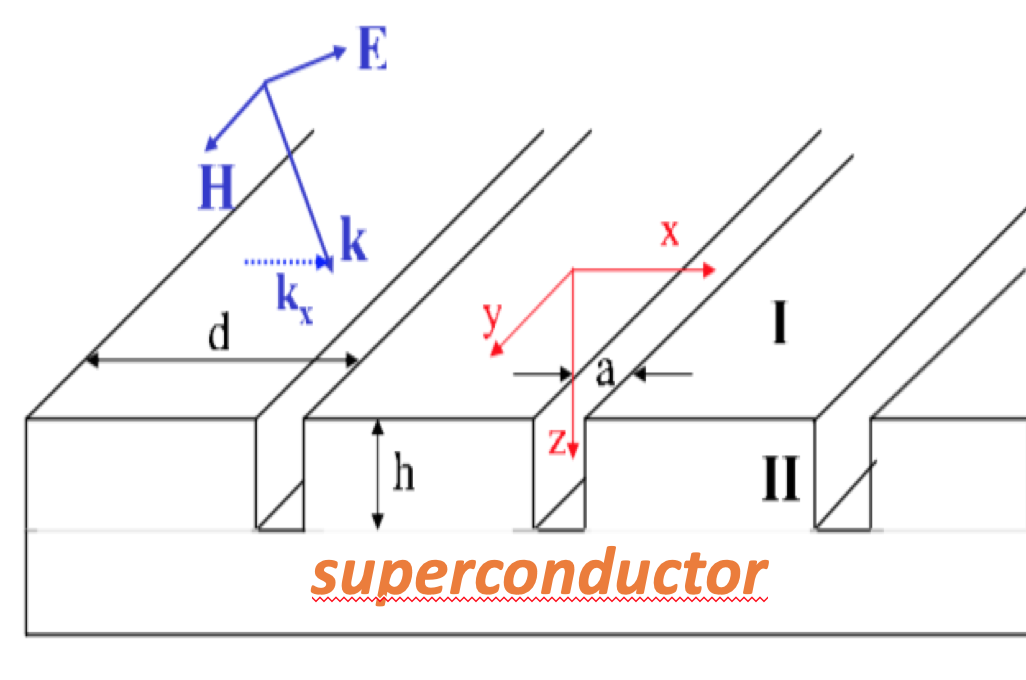}
\caption{Sketch of the periodic structure with grooves digged in the topmost electrode of the  overlap JJ.} 
\label{lioschem2}
\end{figure}

 Infrared radiation impinging from vacuum on the surface of a  semiinfinite normal metal, on the top  of which  an array of grooves  has been drilled  with periodicity $\vec{d} \parallel \hat x$, in the $\hat y $ direction,  of the kind  shown  in Fig.(\ref{lioschem2}),  generates a SPP bound at the surface array and decaying in the inside of the film. The plasma frequency of the SPP dispersion  is dictated by the hole  array size.  In this Section, we recall the simplest derivation of the spoof plasmon bound state\cite{garcia-vidal}.  
 
 Let the top surface be at $z=0$ and the bottom of the grooves be at  $z=h$, so that  the depth of the grooves is $h$ and their width is $a$.  The TE mode of the field,  $E_x, H_y$,  propagating in the  vacuum, can be expressed as the sum of an incident wave and of reflected waves with  reflection coefficients $\rho_n$, where $n$ is the diffraction order.  The subwavelength modulation which provides  diffraction  by the periodic structure,  is unable to resolve the fine structure, so that the radiation  can be averaged in space and  continuously matched  at $z=0$.  
 The $E_x$ field is evanescent  in $\hat z$ ($k_z = i\: \sqrt{ k_x^2-(\omega / c)^2  }$, with $| k_x| > \omega / c$), but,  in the limits  $\lambda >>d>> a $ we can neglect the penetration of  $E_x$   into the semiinfinite bulk of the normal metal  and   impose  its vanishing   at $z= h$. 
As  the wavelength of the radiation is much larger than the  width of the grooves  ($\lambda >2\: n_g\:  a $, where $n_g$ is the refraction index inside the groove), just the fundamental  mode can be considered in the region $-h <z<0$. 
 Within these approximations, a very simple relation arises from the  matching conditions, which provides the   dispersion relation when  reflectivity $\rho_0$ is taken to  diverge\cite{garcia-vidal}:
\bea
\frac{\sqrt{k_x^2-k_0^2} }{k_0} = S_0^2 \: tan(  k_0 h), 
\label{gv}
\enea
where  $\omega  = c \:  k_0 $  and  $S_0^2 =a/d $. At large $k_x$, $ \omega $ saturates at 
$\omega _{spp} = c \frac{\pi}{2 \:h}$, as if the groove acted as a cavity waveguide (vacuum is assumed in the grooves). By choosing $ d=0.45 \: \mu m$, $ h=13 \: \mu m$ and $S_0^2  \sim 0.2$, we find $\omega _{spp} \approx 0.33 \times 10^{14} \: Hz$.  The plot of the SPP,  obtained by solving Eq.(\ref{gv})  is reported in Fig.(\ref{cro})\cite{garcia-vidal}.  The units chosen in the plots  
 for $k_x$ and $\omega$  are $( \pi /d,  \pi c /2h )$.  The additional quasi-linear dispersion appearing  in Fig.(\ref{cro}) is the radiation field due to the fluxon given by Eq.(\ref{plapa}) and discussed in the next Section.

 \section{ fluxon  radiating in a modulated superconducting Josephson Junction }
        
 As discussed in the Introduction, an infrared radiation impinging in free space on the  top electrode of an overlap JJ  of frequency $\omega $  couples  weakly to the fluxon dynamics due to the mismatch between the radiation  wavelength $\lambda $ and the typical length scale - Josephson length - $\lambda_J $ of the fluxon. By modulating the top electrode of an overlap JJ in the form of a MetaMaterial  (MM), sketched  in Fig.(\ref{lioschema}), we find that the interaction between radiation coming from the vacuum and the fluxon can be enhanced.
 
  We consider  a $ S_{MM}/I/S $ long  overlap Josephson Junction of  length $L  \parallel \hat x$. Here $S_{MM}$ stands for one of the superconducting banks, let's say the top one, in which an array of grooves has been drilled in the $\hat y$ direction, as the one sketched in Fig.(\ref{lioschem2}) and presented in Section II. $S$ denotes the bottom uniform and homogeneous superconductor electrode, while $I$ stands for insulator of thickness $d_i$ and width $w$. SPP device based on the proposed layout (see Fig.\ref{lioschema}) could be built using a top-down nanofab techniques that include steps of Electron Beam Lithography, dry and wet etching\cite{dom1,dom2,Goltsman} for writing and then drilling the array of junctions for example, inside a $Nb/NbO_x/Nb$ or $Al/AlO_x/Al$ trilayer sample. We expect that the most relevant effect of the added periodic modulation of the electrode is a corresponding modulation of the Josephson critical current density $j_c$. 
 The inhomogeneities introduced by the diffractive grating attract or repel the fluxon in its propagation. The dips in the modulation tend to attract and localize the fluxon, while the mesas tend to delocalize it. 
 
 The problem was studied long ago both theoretically and experimentally\cite{golubov} in junctions of millimeter size. 
They prove that a fluxon shuttling to and from in a periodically inhomogeneous overlap junction radiates. 
As the derivation of the energy dispersion of the radiating fluxon is based on  perturbation of the fluxon propagating in a homogeneous junction, we start here recalling  the usual approach to the homogeneous problem.  

The gauge invariant form of the supercurrent, written in terms of the phase of the order parameter of the top  and bottom electrodes $\vartheta _{\pm}$ and of the vector potential $\vec{A}$, is:
 \bea
   \vec{ J}^s  =- \frac{2e}{2m} |\psi_0|^2 \left ( \hbar \vec{\nabla}  \vartheta + \frac{ 2e}{c} \vec{A}\right ).
   \label{cura}
   \enea
 Here  $m$ and $-e$, with $e>0$, are the electron mass and charge, respectively and   $ |\psi _0|^2 = n_s $ is the superfluid density. $\vartheta$ is the phase of the superconducting order parameter.  The usual approach to the equation of motion for the phase difference $ \varphi = \vartheta _{+}-\vartheta _{-}$ in a 1-d overlap  junction of  length $L$, along the $\hat x$ axis,   is to consider the  $z$ component of the Maxwell equation:
\bea
\left . \nabla \times B \right |_z-   \frac{\epsilon_r }{c} \frac{\partial E_z}{\partial t }  = \frac{4\pi}{c}\: \left \{ J_J  - \frac{1}{\lambda _J ^2}\:  \frac{V}{R} \right \},\nonumber\\
 \label{usu}
\enea
where $ J_J = J_c \sin \varphi $ is the Josephson current of critical current $J_c$, $V$ is the voltage difference across the barrier and  $R$ is  the quasiparticle  resistance in  the insulating layer.  The length scale characterizing the spacial variation of $\varphi$  is the  Josephson length $\lambda _J$.  To be concrete, estimates will be presented for a junction with Nb superconducting contacts with  $L>> \lambda _J$,  where $ \lambda _J$ is  of the order of  various tens of $\mu m$. The width $d_i$  of the insulating  barrier, along the $\hat z$ axis, is of few $nm$'s.
 
  In the case of bulky superconducting banks one recognises  that   the phase difference  $ \varphi $, at points where the superconducting screening currents  $   \vec{ J}^s$ of  Eq.(\ref{cura})  have vanished, takes the value dictated by unperturbed superconductors.  This allows to relate the Laplacian of $\varphi$ to the  $z$ component of the $curl \: B$  of   Eq.(\ref{usu}), obtaining: 
\bea
\frac{\partial^2 \varphi}{\partial x^2 }+ \frac{\partial^2 \varphi}{\partial y^2 } - \frac{1}{\overline{c}^2}  \frac{\partial^2 \varphi}{\partial t^2 } 
= \frac{1}{\lambda _J ^2}\: \left \{  \sin  \varphi  - \alpha ' \frac{\partial\varphi}{\partial t} \right \}, 
\label{sgor}
\enea
which is the celebrated  Sine-Gordon (S-G) equation for the superconducting phase difference at the overlap junction. Here 
\bea
\overline{c}^2 = \frac{1}{1 + 2 \lambda _L/d_i }\: \frac{c^2}{\epsilon_r}, \:\:\:\:\:\:\: \lambda _J ^2 =\:   \frac{c \phi_o }{8 \pi^2 J_c ( d_i+2\lambda _L)}, \nonumber\\
 \omega _J =\frac{\overline{c}}{\lambda _J } =\left (  \frac{ 2\:e }{\hbar}\frac{I_c}{C}\right )^{1/2},  \:\:\:\alpha '= \frac{\hbar }{2eRI_c}.\: 
\enea
 $\lambda _L$ is the London penetration length, $ \lambda_L ^{-2}   = 4\pi |\psi _0|^2 e^2 / ( m c^2)$ ($\sim 50\: nm$ for $Nb$).  Dimensionally the Josephson critical current  density is $ J_c \sim e/( t {\cal{A}} )$,  where $ {\cal{A}}$  is a cross sectional area pierced by the supercurrent $J$  in the $\hat z$ direction of the overlap junction and t represents the time. We have estimated a $J_C\sim100A/cm^2$ for a device that has  a=d$\sim200$nm and w=$1\mu m$.

 The capacitance  of the junction,  expressed in terms of the thickness of the barrier $d_i$,  $ C = \epsilon _r {\cal{A}}  /({4\pi \: d_i}) $, is rather large, so that charging effects are assumed to be absent.  $\alpha '$ is a  parameter accounting for the ohmic (zero frequency) dissipation. In presence of an incoming radiation of wavelength  $\lambda  \sim 700\: nm$, we have $\lambda_J >> \lambda$. 
 
  In the absence of dissipation ($ \alpha '=0$), the kink  solution for  the  1-d approximation to  Sine-Gordon equation,  Eq.(\ref{sgor}),  is: 
\bea
  \varphi _0( x \pm  u t ) = 4 \arctan \exp \left [\frac{ x \pm u t }{\sqrt{1- u^2 / \overline{c}^2 }} \right ].
  \label{flux}
  \enea
  where $u< \bar c$ is the velocity of the fluxon. 
    
 In the presence of  the perturbation induced by the incoming radiation, an additional field $\vec{B}^{(2)} $ will be considered in  Section V, to be added to the one of Eq.(\ref{usu}). For the time being we consider in this Section only the perturbation induced on the fluxon by the groove array at the top contact. We assume that the effect induced by this modulation is to cause a modulation of $J_c$ :
  $J_c = J_{c0} + J_{c1}\cos \frac{2\pi}{d} x $ in the non dissipative case as follows. If $ J_{c1} < J_{c0}$,  to lowest order, a solution of the motion equation for the  fluxon can be searched by adding  a correction to the unperturbed fluxon of Eq.(\ref{flux}),  as follows: $ \varphi (x,t) =     \varphi_0 (x,t)+  \varphi_1 (x,t) $.   It has been shown\cite{mkrtchyan,golubov} that  the perturbation  $\varphi_1 (x,t) $ can take the form of a plane wave:
 \bea
 \varphi_1 (x,t)  = \sum _{n \neq 0}  A_n \: exp[i( \omega_{pl}  t - k^n_{pl} x)], 
 \enea
  generating a transverse radiation field $\varphi_t  \propto E_z, \varphi_x\propto H_y $  corresponding to the plasma frequency $ \omega^n_{pJ}/2\pi $ and wavevector $  k^n_{pJ}  $ given by  (here  $\nu = \sqrt{1- u^2/{\bar c}^2 }$ and $n$ integer):
 \bea
 \omega^n_{{pJ} ^2} - {\bar c}^2   k^n _{{pJ} ^2} = \omega _J ^2,
  \label{plapa} \\
 \omega^n_{pJ} = \frac{2\pi\: n }{d}\: \frac{u}{ \nu^2}  \pm  \frac{u}{\lambda _J \nu}  \sqrt{ \left ( \frac{ u }{{\bar c}} \frac{ 2\pi \: \lambda _J\: n}{d \:  \nu}\right )^2 -1}
 \nonumber\\
  k ^n_{pJ} = \frac{2\pi\: n }{d}\: \frac{ u^2}{{\bar c}^2} \frac{1}{ \nu^2}  \pm  \frac{1}{\lambda _J \nu} \sqrt{ \left ( \frac{ u }{{\bar c}} \frac{ 2\pi \:  \lambda _J\: n}{d \:  \nu} \right )^2 -1} 
  \label{kpl}
 \enea  
 
 The accelerated  fluxon radiates in the MM and  Eq.(\ref{plapa}) is the dispersion law of the radiation. The approximated form is valid for the far field away from the soliton, with emissions ahead of the fluxon (+), or far away behind the fluxon (-).   It can be shown that the amplitudes $A_n$ of the plasma oscillations decrease exponentially as $n$ increases, so that we will concentrate only on the term $n=1$.  Increasing  $u / {\bar c}$, both $k_{pJ} $ and $\omega_{pJ} $ increase.   An estimate of $k_{pJ} $ for $d = 0.45 \: \mu m$,  $h=13 \: \mu m$, $\lambda _J =100 \:\mu m$,  $\bar c= 0.05\: c$,  $ \omega _J= 1.\times 10^{11} sec^{-1}$  and $u / {\bar c} \sim 0.8$ gives   $ k_{pJ}^-  d/\pi  \sim - 0.89$. The corresponding radiation frequency is, from Eq.(\ref{plapa}), $  \omega _{pJ} \sim 0.93 \times 10^{14} Hz$, which is comparable with the plasma frequency of the SPP.  These parameters are used  in the plot of Fig.(\ref{cro}). Note that $\omega _J $ is about three orders of magnitude smaller than $  \omega _{pJ}$, so that the dispersion of Eq.(\ref{plapa}) is practically linear.  The dependence of $k_{pJ} $ on the fluxon velocity is first order in $ u/\bar c$.  
In the next Section we discuss  a simplified model for the interaction   between the fluxon radiating field and the SPP originated by the MM, which  leads to  absorption of energy from the radiation source. 
 \section{Modes interaction and anticrossing }
 
\begin{figure}
\includegraphics[height=50mm]{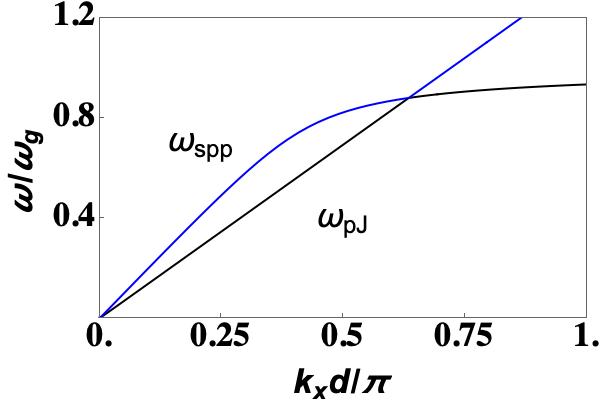}
\caption{Dispersion of the  Spoof Plasmon Polariton,  and  of the fluxon  radiating energy  reported vs $k_x$ on the same plot in unit  $( \pi /d, \omega _g =  \pi c /2h ) $. The 
parameters used are  $d = 0.45 \: \mu m$,  $h=13 \: \mu m$, $\lambda _J =100 \:\mu m$,  $\bar c= 0.05\: c$ and   $ \omega _J= 1.\times 10^{11} sec^{-1}$.} 
   \label{cro}
\end{figure}

\begin{figure}
\includegraphics[height=50mm]{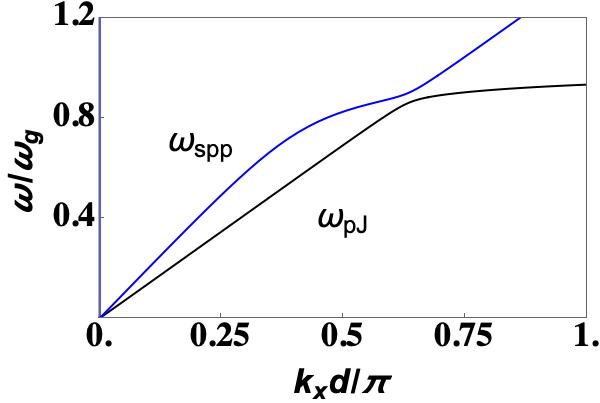}
\caption{Anticrossing at the mode interaction  vs $k_x\equiv k$. The parameters are the same as in Fig.(\ref{cro}). } 
\label{anticro}
\end{figure}

 As shown in Fig.(\ref{cro}), the  SPP dispersion  and the radiation mode of the fluxon  cross  at $ k_x \sim 0.6 \: \pi / d $ for $d=0.45 \: \mu m$ and $h=13 \: \mu m$. The presence or absence of the crossing strongly depends on  the choice of  ratio
 $d/h$.  The fluxon extends over a length $\lambda _J $ much larger than the period of the modulation in the MM, $d$, so that it is reasonable to assume that it moves at an average velocity prior to interaction with a SPP pulse.  The initial velocity should be also determined by accounting for the dissipation mechanisms acting in the dynamics (tuned by the  parameters  $\alpha '$ appearing in Eq.(\ref{sgor}) and $\beta $, to be introduced in the following).  These mechanisms also determine the dynamics of the fluxon and, in turn, its radiative power. We address this point in Section V and Appendices A and B.  In the average,  we assume that the fluxon  keeps an  average  stationary velocity during its motion so that we are in presence of steady state radiation, except when under the action of a short perturbing pulse. This is a very crude approximation, of course, which, however, allows us to modelize  the interaction between the SPP and the fluxon radiation  mode with a very  simple approach. The crossing  in Fig.(\ref{cro}) turns into an anticrossing as shown in Fig.(\ref{anticro}). 
 
 The model rests on few simplified assumptions.
 An electric field from the  MM  in the  non dissipative superconductor  boundary generates a time derivative  of the current density  according to the London equation:
 \bea
  \frac{\partial} {\partial t}  J = \frac{n_s e^2}{m}  E,
  \label{londi}
 \enea
where  $ {n_s e^2}/{m} = c^2/( 4\pi \: \lambda _L^2)$.  The motion equation  (in the $\hat x $ direction along the boundary) for the current in time Fourier transform is ( $   \omega _{pJ} \equiv  \omega _{pJ}( k_{pJ}) $)
\bea 
  - \omega^2 \: J +  \omega _{pJ} ^2 \: J =- i \: \omega\: \frac{c^2}{ 4\pi \: \lambda _L^2} E
  \label{uno}
  \enea
  This is the first equation relating the current density at the boundary with the insulating barrier and the electric field of the radiating fields.
   
\begin{figure}
\includegraphics[height=50mm]{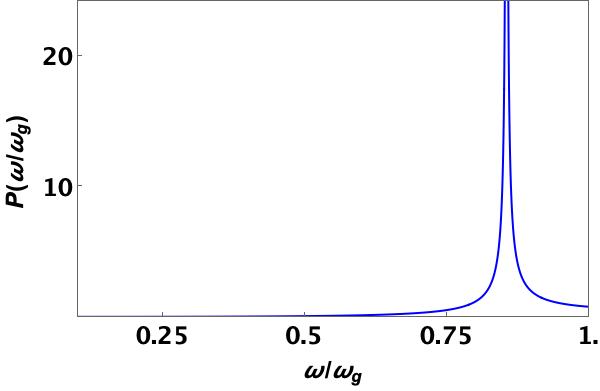}
\caption{Dissipated power at the anticrossing at the mode interaction  vs $\omega$, from Eq.(\ref{dissip}). The parameters are the same as in Fig.(\ref{cro}). } 
\label{pow}
\end{figure}

An additional equation is provided by the relation between the dissipative flow of the added $J$ current and the  electric field. In a normal metal, the resistivity $\rho (\omega ) $ can be related to the dielectric function $\epsilon (\omega ) $ as 
 \bea
(  -i \: \omega  \rho )^{-1} = \epsilon(\omega ) -1.
\enea 
The AC electrodynamics of a superconductor for $\omega < 2 \Delta / \hbar$  ($ \Delta $  is the superconducting gap)  is dominated by the imaginary part of the conductivity, which, at finite temperature, is much greater than the real part in magnitude and is strongly frequency dependent ($\sigma= \sigma _1-i\: \sigma _2 , \: \sigma _2 \sim 1/\omega $).  However, at THz frequencies  ($ > 2 \Delta / \hbar$), the real part of the conductivity plays also a role even at distances from the boundary larger than  $\lambda _L$.  Here we replace $\epsilon $ with  the effective 
  $\epsilon _{xx}$ given by the MM SPP:
  \bea
  \epsilon _{xx}(\omega )  = \frac{ \pi^2d^2 \: \epsilon_g}{8 a^2} \left ( 1 - \frac{ \pi^2c^2 }{\omega ^2  a^2 \: n _g^2} \right ),\nonumber\\
  \omega _{spp}^2 = \frac{ \pi^2c^2 }{ a^2 \: n _g^2} \hspace{3cm}
  \label{oSPP}
  \enea 
  where $\epsilon_g$ and $n_g$ are the dielectric constant  and refraction index of the material in the holes\cite{pendry}.  $ \omega _{spp}$  is assumed to be rather independent of $k_x$ in the range where the dispersion has reached saturation.
   In this approximation,  
  the motion equation for the electric field at the boundary is : 
  \bea
 - \omega ^2  E   + \omega_{spp}  ^2  E  = -\omega ^2 \:  \rho(\omega ) \:J (k_x,\omega ).
 \label{due}
 \enea
 The system of Eq.s (\ref{uno},\ref{due}) provides the eigenvalues which are solution of:
 \bea
 \omega ^4  -\omega ^2  \left ( \omega_{spp}  ^2 +  \omega _{pJ} ^2 +  i \:  \frac{  \omega \:c^2}{ 4\pi \: \lambda _L^2} \: \rho (\omega )  \right ) +  \omega _{spp} ^2 \omega _{pJ} ^2  =0\nonumber
 \enea
The anticrossing which arises from this very crude approach appears in  Fig(\ref{anticro}).
The dissipated power can be extracted by  squaring  Eq.(\ref{due}) and by  using $ \rho(\omega ) = [- i \omega  \left ( \epsilon _{xx}(\omega)  -1\right )] ^{-1} $:
\bea
- i \: \omega |E|^2 \propto  \frac{ 1}{\left ( 1- \frac{\omega _{spp}^2(k_x)}{\omega ^2} \right )^2}  \frac{1}{  \left ( \epsilon_{xx}(\omega) -1\right ) } \:  \rho (\omega ) |J|^2
\label{dissip}
\enea 
and has been plotted  in Fig.(\ref{pow}) as a function of $\omega$. It is peaked at the crossing point,  $k_x d/\pi \sim 0.6$.
 Eq.s(\ref{uno},\ref{due})  are  coupled to the equation of motion of the fluxon, Eq.(\ref{sgor}),  because  the propagation velocity $u$ is required to define the dispersion of the radiating field generated by the fluxon, Eq.(\ref{plapa}). Forcing terms acting in the superconducting phase dynamics arising from $B_y^{source}$, the magnetic field generated by  the TEM incoming wave,  are derived in the next Section. 
 
 \vspace{0.5cm}

  \section{ The  fluxon motion equation in presence of radiation}     
  
The goal of this Section  is to extend the S-G equation, Eq.(\ref{sgor}),  to include the presence of the SPP perturbation and  the induced radiating fields.  Indeed, the modulation in the top contact due to the presence of the grooves is responsible for extra radiation by the fluxon during its dynamics. Special concern, in presence of radiating fields, is for the dissipation mechanisms in the junction. Both effects generate a current imbalance at the interface $J_+-J_-$, and  an added  extra field $ B^{(2)}$.  Extension of Eq.(\ref{usu}) implies that two extra terms have to appear in  Eq.(\ref{usu}):  $ curl\: B ^{(2)} $ in the first term on the l.h.s. of  Eq.(\ref{usu}) and the imbalance current  $J_+-J_-$ on the r.h.s.  of Eq.(\ref{usu}). They will account both for the SPP generated by the incoming radiation and for the radiating fluxon itself. As discussed in  Appendix C,  Eq.(\ref{cura}) has to be rewritten as (Eq.(\ref{ino2})):
\bea 
\frac{\phi _0}{2\pi }\:   \frac{\partial \varphi }{\partial x } =  \frac{4\pi  \lambda_L^2} {c }  \left [ J_+-  J_- \right ]- ({ 2\lambda _L +d_i})\: B_y  -  \lambda _p \: B_y ^{(2)}. \nonumber\\
    \label{inok}
    \end{eqnarray} 
 The penetration length $\lambda _p$ is discussed here below.

  In Appendix C we derive an expression for  $ curl\: B ^{(2)} $  which contributes  to the motion equation of  Eq.(\ref{sgor}) with a dissipative term, the usually  called ' $\beta-$term' ( see Eq.(\ref{difg}) given here below).
   Finally, using  the London equation  (see  Appendix C, Eq.(\ref{Ctota})) we obtain \cite{bulaevskii}:
\bea
\left . \nabla \times \nabla \times \vec{B}^{(2)} \right |_y + \frac{1}{ \lambda_\omega  ^2}  \: B_y^{(2)} =-  \frac{4\pi}{c} \frac{ \hbar }{2e\: d} \: \sigma _{qp}\frac{\partial^2 \varphi  }{\partial t\partial x }, 
\label{rota6}\\
{\rm with}\:\:\:  \frac{ 1}{ \lambda_\omega  ^2}   \equiv  \frac{1}{ \lambda_L ^2} +  \epsilon _{b}(\omega )  \frac{\omega ^2}{c^2} \hspace{3cm}
     \label{lamom}\\
  {\rm and }\:\:\:\:\:\:  \epsilon_b (\omega )  = \epsilon_b +   \frac{4\pi \: i \: \sigma _{b} }{c^2\: \omega}. \hspace{3cm} \nonumber
 \enea

 Eq.(\ref{lamom}) can be interpreted as follows.  $2 \pi /\lambda_\omega $ plays the role of  $ k_x$. By choosing  $ \epsilon_b \sim 41.4 $ ( Nb in the THz range  $ \omega \sim \pi \times 10^{13} Hz$),  we get  $\lambda _\omega ^{-1}  \sim  2.1 \times 10^6 m^{-1} = k_x/2\pi $ ($\lambda _L \sim 50\:  nm$). On the other hand,
  \bea
   -ik_z = \sqrt{k_x^2 - \omega ^2/c^2} \sim\frac{ \pi}{d}  \sqrt{(0.63)^2- \left (\frac{0.8}{2} \frac{d}{h}\right )^2 }\nonumber\\
   \sim  4.4 \times 10^6 m^{-1}.
   \enea
    $k_z$  is purely  imaginary and provides the decay of the $B^{(2)}$ field within the  top superconducting contact. We define the inverse of $|k_z|$, as the penetration depth  $\lambda _p $ of the field $\sim 230 \: nm$. 

  Eq.(\ref{rota6}) with the definition of $\lambda_\omega $ of Eq.(\ref{lamom}) are the starting point of our analysis. 
With   an exponentially decaying dependence on the $z$ coordinate, $ e^{-z/\lambda_p}$ of all the fields involved within the overlap junction region we can introduce an effective 1-d  Green function $  G(x , x',\omega ) $, with zero boundary conditions at $x=0$ and $x=L$. $  G(x , x',\omega ) $ inverts the differential operator in 
   Eq.(\ref{rota6}) with $i\: k = \left [ 1/ \lambda _\omega^2 - 1 /\lambda _p^2\right ]^{1/2}$ ( $k$ real):
    \bea
  \left [   \vec{\nabla }^2 - k^2 \right ]\: G\left (x,x',\omega \right ) = -\delta  \left  (x -x'\right ).
  \label{rota5}
  \enea 

Hence, $ \nabla _x B_y^{(2)} ( x,t ) = \left. \nabla \times \vec{B}^{(2)}\right |_z$,  solves the integral equation
 \begin{widetext}
\bea
 \nabla _x B_y^{(2)} ( x,t ) = \nabla _x B_y^{source} (x,t )  -\frac{ 1}{\lambda_L ^2 }  \int dx'  \:dt'   \:\nabla _x \:  G(x ,x', t-t') \:  {\cal{F}}\left [ \varphi (x', t' ) \right ] . 
 \label{roro} \label{difg}
 \enea
  \end{widetext}
  where 
  \bea 
    {\cal{F}}\left [ \varphi (x,t) \right ]  =  \frac{2\:\phi_o\:  \lambda_L ^2}{d\:c^2}\: \sigma _{qp}\:\frac{\partial^2 \varphi  }{\partial t\partial x }
    \label{lam} 
    \end{eqnarray}
and we have also added  the contribution of $ B_y^{source}$ as an inhomogeneous term.  to be included in  Eq.(\ref{usu}). Deriving  Eq.(\ref{inok}) with respect to $x$, we insert  $  \nabla _x B_y^{(2)} (x,t) $ from Eq.(\ref{roro}) in it  and   divide the resulting equation    by $J_c /c $.  The extended  form of Eq.(\ref{sgor}), which includes a $\beta-$ dissipative term  is obtained:

 \begin{widetext}
  \bea
{\lambda _J ^2}\: \frac{\partial^2 \varphi}{\partial x^2 }-  \frac{1}{\omega _J^2} \: \frac{\partial^2 \varphi}{\partial t^2 }- \alpha  \:\frac{1 }{\omega _J} \: \frac{\partial \varphi}{\partial t } + \beta  \int  dt'  \int  dx' dz' \: G(x,x',t-t') \: 
\frac{\partial^3 \varphi  }{\partial t\partial x\partial x } - \sin [\varphi (x,t) ] +\gamma  \nonumber\\
=  \frac{8\pi^2  \lambda_L^2} {\phi _0 c }  \nabla _x  \left [ J_+-  J_- \right ] + \frac{\lambda_p}{d_i+ 2 \lambda _L}  \frac{c}{J_c}\nabla _x B_y^{source} (x,t) , \hspace{4cm}
  \label{roro3}\\
\alpha = \frac{\hbar \omega _J }{2eRI_C}, \: 
 \:\:\: \: \beta = \sigma _{qp} \:  \frac{2\phi_0 \: \omega _J}{ J_c\: c\: d_i } ,\phi_0=hc/2e,  \:\:\: \gamma = \frac{J_{ext}}{J_c}. \hspace{1cm}\nonumber
 \enea
\end{widetext}

We  have integrated  the  integral of Eq.(\ref{difg}) by parts. The term at the boundary vanishes due to the chosen boundary conditions, so that the '$\beta-$term' displays the third order derivative of the field $\varphi$. A current source term $\gamma$ has also been included.  

Usually  no retardation is assumed in Eq.(\ref{roro3}),  so that $  G({x} ,{x'},\omega ) \sim G_k(x ,x'))$. The Green's  function could account for the periodicity of the grooves potential following the lines of Ref.\onlinecite{golubov} but, as  $\lambda _J >> d$, we can expect that the modulation of the potential is on a much smaller scale than the scale characterising the fluxon dissipation so that we can treat the superconductor MM as an effective homogeneous medium. This is  consistent with a similar approximation which gives rise to the SPP dispersion. Moreover, it is customary to turn to a local approximation for the kernel, so that  the integral in Eq.(\ref{roro3}) disappears. Then, the motion equation for the phase driven by the plasmonic magnetic field  $B_y^{source} $, in dimensionless coordinates, $ t \to \omega _Jt$ and $x \to x/ \lambda _J$, takes the usual form (see below): 
 \bea
\varphi_{xx} - \varphi_{tt} -\sin \varphi = \alpha \: \varphi_t - \beta \: \varphi_{xxt} -\gamma  -   g(t,x) 
\label{motion}
\enea
where $g(t,x) $ includes the forcing terms, on the r.h.s. of Eq.(\ref{roro3}).

We concentrate now on the two added terms included in $g(t,x)$, produced by the SPP. From Eq.(\ref{uno}), the Fourier transform of the current difference term gives: 
    \bea
   - \frac{4\pi  \lambda_L^2} {c }   \left  [ J_+-  J_- \right ]  =  i \: \frac{ \omega }{\omega ^2 -\omega ^2_{pJ}} \:c \:\delta  E_x,
    \enea
    where $\delta  E_x$ is the difference in electric field component between the upper and lower boundary of the junction.  
    Similarly, from $ \vec{k}\times \vec{E} = \vec{B}\:  \omega /c$, the last term reads: 
    \bea
    -  \lambda _p \: B_y ^{source} =-i \: \frac{c |k_z |\lambda _p}{\omega } \: \delta E_x.
    \enea
    where $-i \: k_z = \sqrt{ k_x^2-(\omega / c)^2  } >0$ (here $\omega $ is the frequency of the source radiation).    
    We get: 
    \bea 
    g(\omega, x )= - i \frac{2e \lambda _J^2 }{\hbar \omega _J}\left [\ \frac{ \omega }{\omega ^2 -\omega ^2_{pJ}} 
    - | k_z|\lambda_p \frac{4 \pi }{  \omega } \right ] \: \nabla_x  \delta E_x(x).
    \label{perturb}
    \enea
 At $\omega \approx \omega _{spp}$ the charge density modulation induced by the SPP, $ \rho _{spp}$, appears, as $ \frac{\partial}{\partial x}   \delta E_x^{spp}(x) = 4 \pi \rho _{spp}(x)$. Here $Q_{spp} \equiv   \lambda _p w L \: \rho _{spp} $ is defined as the charge imbalance induced by the oscillating SPP. We average over the length $\lambda _J $ in the $\hat x$ direction, assuming an oscillating dependence $ e^{ik x}$ and in the transverse directions of cross-section $ \lambda _p w$. We rewrite Eq.(\ref{perturb}) in terms of the amount of charge, $Q_{spp}$,  singling out just one frequency $\omega \approx \omega _{spp}$:

\bea
g(\omega, x,t) \sim   \frac{4\pi }{\hbar \omega _J} \frac{\omega }{\omega ^2-\omega _{pJ}^2}  \frac{\lambda _J}{w} \frac{\sin{ k \lambda _J}}{k L}\: \frac{e  \:Q_{spp} }{\lambda _p} \: cos \: \omega  t.\nonumber\\
\label{perturb2}
\enea 
 The largest contribution to the perturbation comes from the first term  of Eq.(\ref{perturb}), at  frequency $\omega \approx \omega _{spp}\approx \omega _{pJ}$. 
When the fluxon velocity  provides a $k_{pJ}$  close to the point at which the two dispersions cross, (see  Eq.(\ref{kpl})),  the perturbation enters a resonance with the excitation modes  of the combined system and its effect is largest. According to our parameters and to Fig.(\ref{cro}), this occurs at velocity   $u = 0.6 \bar c$ which corresponds to  $ k_{pJ} d/\pi \sim 0.7$.
\section{Dissipationless simulated dynamics of the perturbed fluxon } 
\begin{figure}
  \centering
\def\big{\includegraphics[height=5cm]{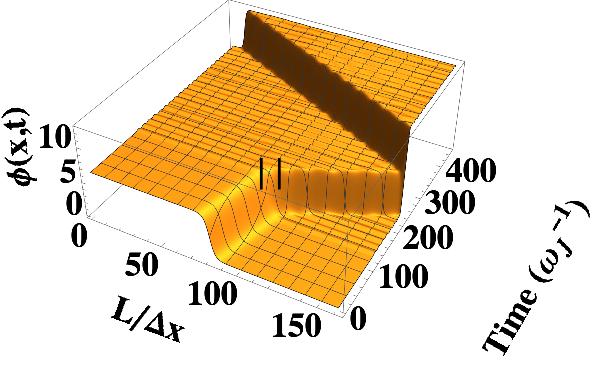}}
\def\little{\includegraphics[height=2.3cm]{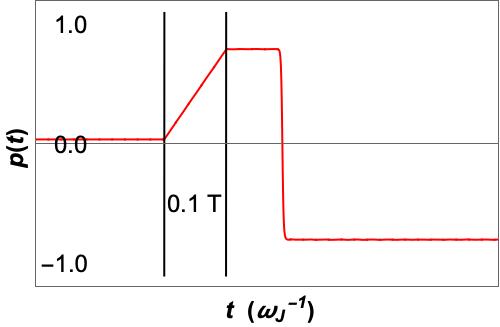}}
\def\stackalignment{l}
\topinset{\little}{\big}{-25pt}{-12pt}
\caption{  3D-plot of the fluxon amplitude $\phi$ vs $x$ and $t$.  The initial velocity is $u/ \bar c=0.1$, far from resonance. A square pulse of small amplitude ( $A=- 0.1$,  see text) acts between times 187 and 220 ( duration $0.1\: T$), marked by the black slashes. The fluxon is speeded up by the pulse. The inset shows the impulse of the field  $P(t)$ vs $t$ according to eq.(\ref{impo}). The pulse acts in the time interval marked by the black lines.} 
\label{gigm01}
\end{figure}

\begin{figure}
  \centering
\def\big{\includegraphics[height=5cm]{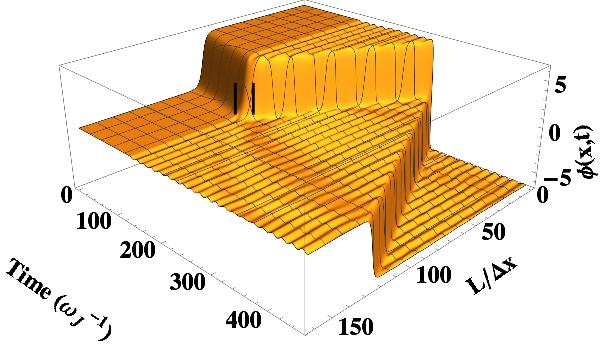}}
\def\little{\includegraphics[height=2.3cm]{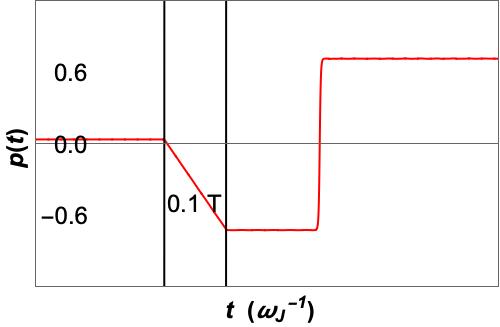}}
\def\stackalignment{l}
\topinset{\little}{\big}{-25pt}{-12pt}
\caption{  3D-plot of the fluxon amplitude $\phi$ vs $x$ and $t$.  The initial velocity is $u/ \bar c=0.1$, same as Fig.(\ref{gigm01}). The amplitude of the square pulse is $A= 0.1$. The fluxon is scattered backward by the pulse. The inset shows the impulse of the field $P(t)$  vs $t$. } 
\label{gig01}
\end{figure}

\begin{figure}
  \centering
\def\big{\includegraphics[height=5cm]{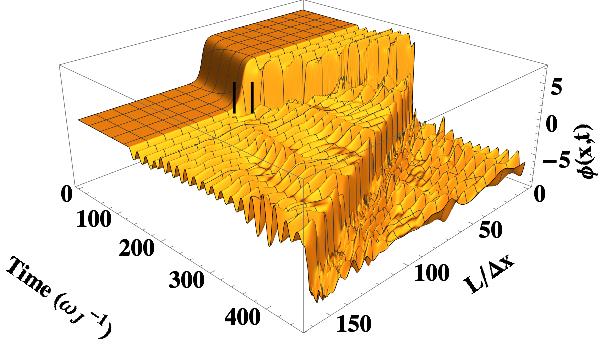}}
\def\little{\includegraphics[height=2.3cm]{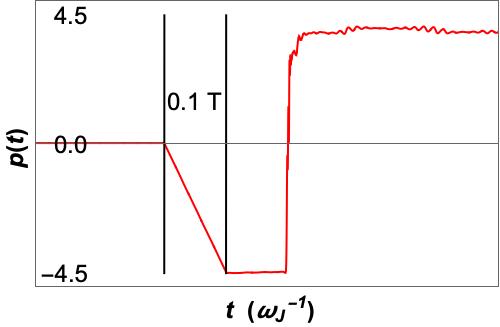}}
\def\stackalignment{l}
\topinset{\little}{\big}{-25pt}{-12pt}
\caption{  3D-plot of the fluxon amplitude $\phi$ vs $x$ and $t$,  same as Fig.(\ref{gigm01}), for  initial velocity  $u/ \bar c=0.1$  and  $A= 0.6$. The fluxon is speeded up, but its shape is conserved except for beatings which mark the approach of a critical perturbation. The inset shows the impulse of the field $P(t)$  vs $t$. } 
\label{gig06}
\end{figure}

\begin{figure}
  \centering
\def\big{\includegraphics[height=5cm]{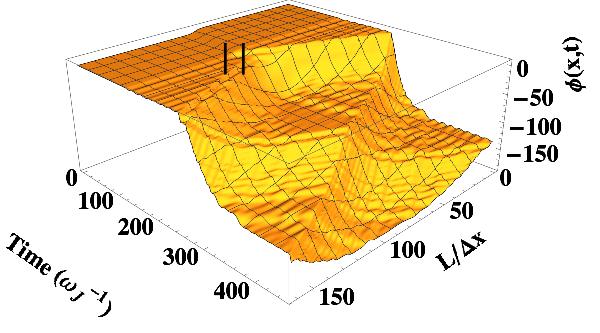}}
\def\little{\includegraphics[height=2.3cm]{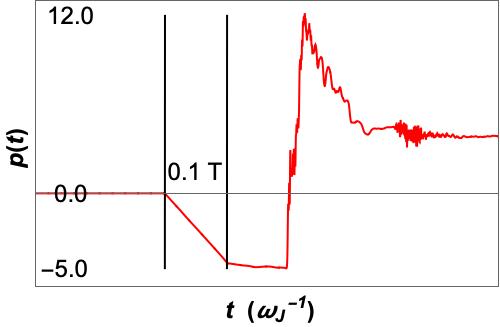}}
\def\stackalignment{l}
\topinset{\little}{\big}{-25pt}{-20pt}
\caption{  3D-plot of the fluxon amplitude $\phi$ vs $x$ and $t$,  same as Fig.(\ref{gigm01}), for  initial velocity  $u/ \bar c=0.1$  and  $A= 0.62$.  The perturbation scatters the fluxon both backwards and forwards. The various components of the field interfere heavily and the  fluxon  itself  is lost,  while acquiring and eventually loosing  extra impulse (see inset). The phase difference rolls down with time when the effect of the pulse ( but not the acquired impulse) is over.} 
\label{gig062}
\end{figure}

\begin{figure}
  \centering
\def\big{\includegraphics[height=5cm]{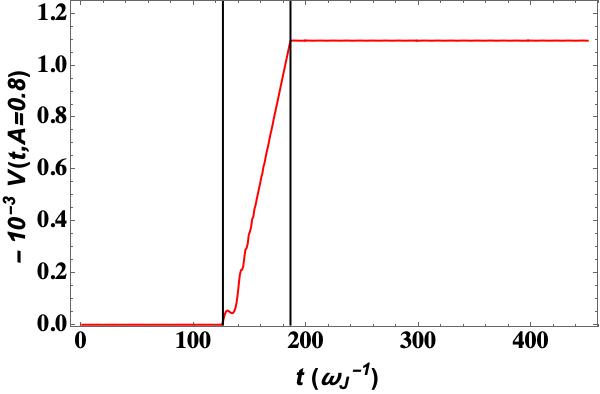}}
\def\little{\includegraphics[height=3.0cm]{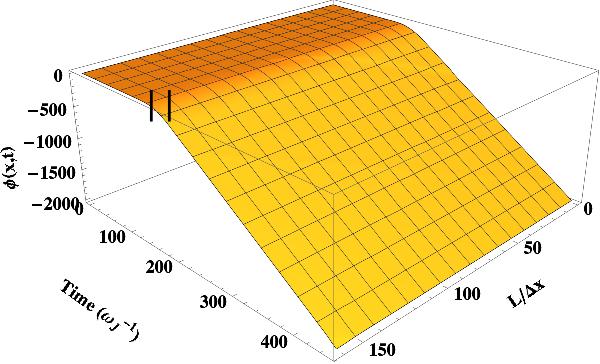}}
\def\stackalignment{r}
\bottominset{\little}{\big}{30pt}{-29pt}
\caption{ Integrated  Voltage in units $ \hbar \omega _J/2e$ vs  $ t $  for  initial velocity  $u/ \bar c=0.1$  and  $A= 0.8$ ( see Eq.(\ref{vovo}). The phase 
$\phi$ vs $x$ and $t$ rolls down almost uniformly in time  for any $x$  as shown in the  3D-plot inset.The pulse acts in the time interval marked by the black lines. } 
\label{vov08}
\end{figure}

\begin{figure}
\includegraphics[height=55mm]{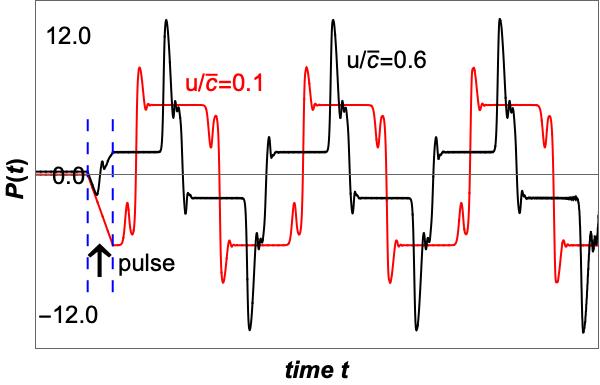}
\caption{Long time comparison of impulse  $P(t)$ vs  $t$ at perturbation amplitude $A= 0.8$  between  $u/\bar{ c}=0.1$ and  $u/\bar{ c}=0.6$.  The periodicity arises from reflection at the boundaries.  The step of the  time scale has been adjusted to  present graphically comparable periods for the two velocities. The arrow points at the time interval during which the square pulse  is active.} 
\label{lu0106}
\end{figure}

\begin{figure}
  \centering
\def\big{\includegraphics[height=4.8cm]{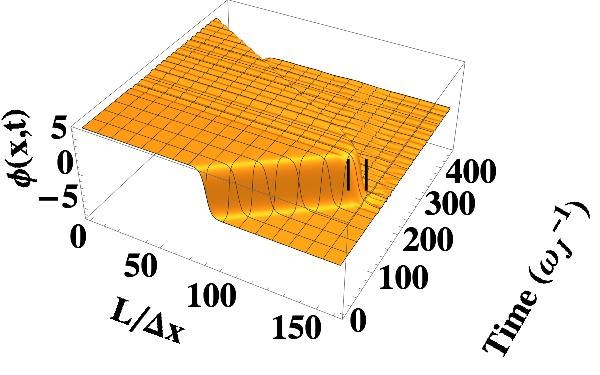}}
\def\little{\includegraphics[height=2.2cm]{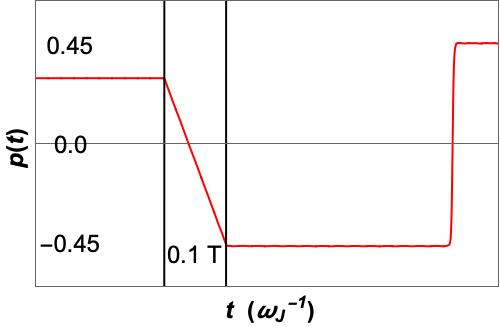}}
\def\stackalignment{l}
\topinset{\little}{\big}{-25pt}{-12pt}
\caption{  3D-plot of the fluxon amplitude $\phi$ vs $x$ and $t$.  The initial velocity is $u/ \bar c=0.6$, which locates the $k$ vector close to the resonance, according to Eq(\ref{kpl}).A square pulse of small amplitude $A= 0.1$,  acts between times 187 and 220 (duration $0.1\: T$), marked by the black slashes. The fluxon is backscattered by the pulse. The inset shows the impulse of the field $P(t)$  vs $t$ according to eq.(\ref{impo}). The pulse acts in the time interval marked by the black lines.} 
\label{rig01}
\end{figure}

\begin{figure}
  \centering
\def\big{\includegraphics[height=4.8cm]{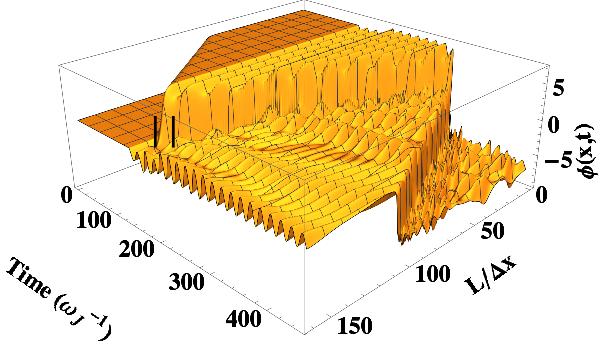}}
\def\little{\includegraphics[height=2.2cm]{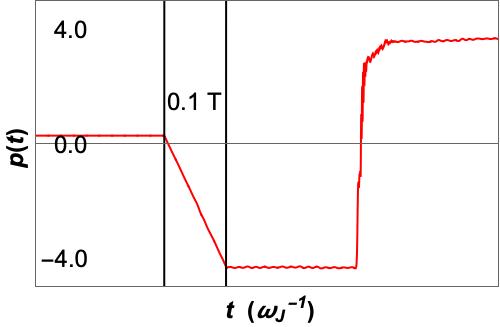}}
\def\stackalignment{r}
\topinset{\little}{\big}{-50pt}{10pt}
\caption{  3D-plot of the fluxon amplitude $\phi$ vs $x$ and $t$.  The initial velocity is $u/ \bar c=0.6$ (same as Fig.(\ref{rig01})), with a pulse amplitude  $A= 0.6$, close to the critical value. Heavy beating form but the fluxon shape can still be recognized. The inset shows the impulse of the field $P(t)$  vs $t$. } 
\label{rig06}
\end{figure}

\begin{figure}
  \centering
\def\big{\includegraphics[height=5cm]{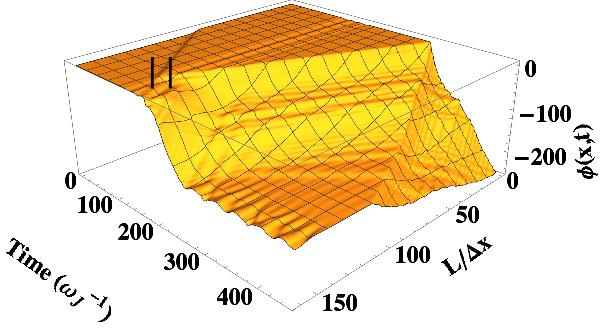}}
\def\little{\includegraphics[height=2.3cm]{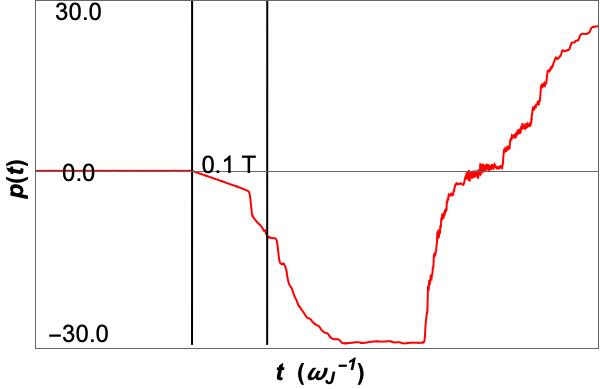}}
\def\stackalignment{r}
\topinset{\little}{\big}{-48pt}{16pt}
\caption{3D-plot of the fluxon amplitude $\phi$ vs $x$ and $t$, same as Fig.(\ref{gigm01}), for initial velocity $u/ \bar c=0.1$ and $A= 0.6$. The fluxon is speeded up, but its shape is conserved except for beatings which mark the approach of a critical perturbation. The inset shows the impulse of the field   $P(t)$  vs $t$. } 
\label{rig065}
\end{figure}

Let us now consider a dissipationless dynamics of the fluxon perturbed at some given  time $t_0 >0$ by a short square pulse, acting  for a restricted time interval $\sim 0.1 \: T$. The effect of the perturbation depends on the incoming velocity of the fluxon, and, of course, on the perturbation strength. Depending on its sign, the perturbation can increase or decrease the  propagation velocity of the travelling  fluxon, and can even scatter back the fluxon. The sequence of figures Fig.(\ref{gigm01}-\ref{rig065}) shows  the 3-d plots of  the simulated dynamics of the fluxon $\varphi ( x,t) $ vs $x$ and $t$, in units $\lambda_J$ and $ \omega _J$.  The maximum displayed time in these plots is $T = 450 \omega _J ^{-1} $, while the length of the junction is $L \sim 25 \: \lambda _J$. In the time interval $\Delta t \in ( 187,220) \sim 0.1  \: T $ a square pulse  of the form  $ A \: cos \: \omega _{pJ} t $, of amplitude $A$  is turned on, with $ \omega_{pJ}/ \omega _J= 0.33 \times 10^{3}$.
The first sequence of plots, Fig.(\ref{gigm01} - \ref{gig062}), monitors  the propagation of a fluxon of   incoming velocity   $ u / \bar c=0.1$.  For $A=\pm 0.1$ the fluxon is just speeded up ( Fig.(\ref{gigm01})) or slowed down till to velocity inversion ( Fig.(\ref{gig01})), respectively. The insets show the change in impulse $P(t,A) $ as a function of time: 
\bea
P(t;A) \propto \int _0^L dx\: \varphi_t(x,t;A)\varphi _x(x,t;A)
\label{impo} 
\enea
with flip of sign when the fluxon hits the junction  edge and is reflected. During the time of the pulse the impulse increases approximately linearly and stabilizes at a higher value, when the perturbation is turned off. From the pulse switch off time, onward, some beating can be seen in the fluxon amplitude time dependence, which is  left over by the perturbation. The fluxon is reflected when it reaches $x=L=160$, as seen from the change of slope of the field and from the sudden change of sign of  the impulse  in the inset.  Since then, the motion is again at constant impulse, but backwards. At the reflection, the amplitude of the fluxon  jumps by $2\pi$.

 It is noticeable that the various scatterings induced by the pulse, with coexistence of forward and backward propagating waves,  end up in an impulse which  is strictly  periodic  with the dwelling of the superconducting phase excitation inside the Josephson junction. This is due to the fact that dissipative terms have not included in the dynamics. 
The overshooting at each reflection is clearly seen.  


By increasing the amplitude, A, of the forcing term, there is no
qualitative change in the time evolution of the fluxon, till A reaches the
value A $\sim$ 0.6 (Fig.(\ref{gig06})). Subharmonic oscillations and beating
markedly increase but the amplitude of the kink is still limited to the
2$\pi$ flux jump. Beatings appear as a consequence of screening of the
incoming kink by the collection of scattered antikinks as it happens
when an electric charged particle is screened by a bath of opposite
charges. This is the classical analogue of Friedel oscillations appearing
in quantum scattering.

Fig.(\ref{gig062})  shows the $\varphi$ evolution for  $A=0.62$. The pulse acts as a strong scattering potential, so that there are scattered components of the original fluxon which move backward and forward with different velocities, generated by the pulse itself which are reflected at the boundary $x=0, L $. The various components of the field interfere heavily and the fluxon kink  is fully lost.  Even when the effect of the pulse is over, such an interference generates  an overshooting in the impulse at the reflection  at the boundary, which is reabsorbed in a finite time in the multiple interference processes (see inset). Again, the inset shows the impulse of the system as a function of time. However, it is a space integrated quantity, so that it captures only the average of the complex evolution. During the overshooting time and beyond, the phase field rolls down to higher and higher values, as confirmed by the voltage difference at the junction integrated over the whole junction length: 
\bea
V(t;A) \propto  \int_0^L dx\: \varphi_t (x,t;A),
\label{vovo}
\enea
which is plotted in Fig.(\ref{vov08}) for $A=0.8$. Of course, the physical phase difference amplitude is  $mod[2\pi]$. While the shape of the kink is lost, the propagation across the junction with reflection at the edges survives and strongly characterizes   the impulse for larger evolution in times. This is reported  in Fig.(\ref{lu0106}) for $A= 0.8$ ({\sl red curve  for } $u/\bar c= 0.1$).

The sequence Fig.(\ref{rig01}-\ref{rig065}) corresponds to the  sequence  Fig.(\ref{gigm01} - \ref{gig062}) but for  $u/\bar c= 0.6$, which implies that the $k_x$ vector is close to the point where anticrossing occurs in Fig.(\ref{anticro}). We made the choice of keeping similar strength of the perturbation $A$, although, according to Eq.(\ref{perturb2}), there would be an extra  factor of $\sim 20$ included in  $A$ in this  case, to account for the vicinity to  the pole.  Apart for the obvious increase of the strength of the perturbation due to this extra factor, two features can be noticed when the initial velocity of the fluxon increases. Comparing Fig.(\ref{gig06}) with Fig.(\ref{rig06}) with same perturbation strength $A$, the evolution of the fluxon having initial velocity $u/\bar c= 0.6$ appears to be less sensitive to beating and subharmonic oscillations than when the fluxon is moving slower. On the other hand the overshooting of the impulse when the fluxon inverts its motion at the edges is even larger as shown in Fig.(\ref{lu0106}). This corresponds to a faster roll down of the phase as marked by the larger scale for $\varphi $, which appears in  Fig.(\ref{rig065}), when compared to  Fig.(\ref{gig06}).

Let us now inquire up to what SPP charge $ \tilde Q_{spp}$,  the fluxon may be assumed to be  insensitive to the pulse. In other words, to which extent the forcing term can simply neutralize some dissipation induced by a term $ -\alpha  \varphi _t$ appearing in the motion equation.  Let $\tilde T$ be the time scale of the SPP pulse. Qualitatively, ignoring the $\beta-$term which is expected to be small, we can estimate a compensation in the average, between the forcing term and the dissipative  $\alpha -$term: 
\begin{widetext}
\bea
-\alpha \int_{-L/2}^{+L/2} dx\int_0^{\tilde T}  dt  \: \varphi _t \: e^{-t/{\tilde T}} \cos \omega_{pJ}  t = 8 \pi ^2 \rho _{SPP} \frac{w}{\phi _0}   \int_{-L/2}^{+L/2} dx\int_0^{\tilde T}  dt  \: e^{-t/{\tilde T}} \cos^2 \omega_{pJ}   t , \nonumber\\
i.e. \hspace{1cm}
2 \pi \alpha \frac{u}{1 + (\omega_{pJ}  {\tilde T})^2 } =  \lambda _p L w \: \rho_{spp} \: \frac{\lambda _J}{\lambda _p} \: \frac{ 1+2 (\omega_{pJ} {\tilde T})^2}{1+4 (\omega_{pJ} {\tilde T})^2},  \hspace{7cm}
\enea
\end{widetext}
where the unperturbed  fluxon waveform $\varphi (x-ut) $ of Eq.(\ref{flux}) has been used.

  The requirement that $u < 1 $ implies ( $ \omega _J {\tilde T} << 1 <<  \omega_{pJ}  {\tilde T}$) that  the overall induced charge by the SPP,  $ \tilde Q_{spp} =  \lambda _p L w \: \tilde \rho_{spp} $, has to satisfy the inequality:
\beq
  \tilde Q_{spp} < \frac{\alpha }{2\pi}  \frac{\lambda _p}{\lambda _J} \frac{1}{ (\omega _{pJ} {\tilde T})^2},
  \eneq
which is quite a stringent condition on the intensity  of the incoming radiation.

  \section{Conclusions}
  
  	Integrating superconductive and optical networks in a low temperature environment is becoming more and more desirable for quantum information processing, but it faces a longstanding problem. While optical fibers and optical circuits mostly involve frequencies in the infrared or, recently, THz frequency window, typical frequencies of a superconducting device are in the microwave range. On the other hand the possibility of putting  fluxons travelling in a long Josephson  Junction (JJ) in interaction with optical signals would increase enormously their flexibility as a tool for  biasing  and controlling gates in a classical or quantum circuit. 
	Recently optically generated Spoof Plasmon Polaritons (SPPs) can be read out by means of integrated superconducting single-photon detectors\cite{heeres} and, in general, interaction of a  Josephson Junction   with  a surface plasmon allows to limit the power delivered to the junction and to  avoid large increase of quasiparticle excitations. Still, optimization of energy exchange between a surface plasmon and a fluxon requires that the difference in frequency between the two-excitation modes is somehow reduced.  We have shown that a feasible way to reach this goal is to engineer one of the banks of the JJ in the form of a metamaterial (MM) which has been proved to generate a SPP  at THz frequency\cite{garcia-vidal}. The SPP can be absorbed by the moving fluxon. 
	
	We have shown that  the two excitation modes, SPP and fluxon radiative field,  can interact  (see Fig.(\ref{anticro})).  Indeed, the MM bank induces a radiative field by the fluxon  of comparable frequency. The typical anticrossing in the dispersion is due to charge oscillations at the MM bank, which gives, rise to absorption of impulse by the fluxon. The latter can be speeded up or slowed down or even scattered backwards by   interaction with a pulsed  SPP, which acts as a forcing term  on the Sine-Gordon (S-G) dynamics of the fluxon, driven by the oscillations of the SPP charge. 
	
	  We provide examples of the simulated S-G dissipationless dynamics of a fluxon in a long  JJ  in which a  free propagating fluxon is acted on  by the SPP perturbation for a limited time interval. The boundary conditions for $\varphi _x$ in the motion equation are standard\cite{lomdahl}.
We show that the fluxon field acquires subharmonic oscillations and beating, due to  extra impulse  absorbed from the perturbation, without loosing its  kink shape, unless the perturbation amplitude is higher than a critical value, which depends on the initial velocity, that is on the vicinity to the anticrossing point. Indeed the fluxon keeps being rather robust  but, meanwhile, it  generates interference of  extra $2\pi-$jumps. This happens because the soliton energy is two orders of magnitude larger than the energies of the excitation modes of Fig.(\ref{anticro}), involved. For perturbation amplitudes higher  than the critical value, the fluxon field looses its shape, but not the periodic dwelling motion, with reflections at the edge of the junction. The superconducting phase rolls down as in a washboard potential and a marked kink in the voltage appears (see  Fig.(\ref{lu0106})). Increasing the initial velocity of the fluxon makes it more robust up to the critical perturbation strength, but lowers the threshold of criticality quite a lot, because the anticrossing point is approached. 
	  
	   We argue  in Appendix B, with an approach similar to the standard one reported in Appendix A,  that dissipative terms in the motion equation do not affect qualitatively the fluxon motion provided an applied current bias is fed in the junction. This is because the MM has a structure on a scale of hundreds of $nm$, much smaller than the typical length scale of the junction dynamics $\lambda _J$. However this also requires that the junction itself is quite long, up to millimeters.
	   
	   \vspace{0.5cm}
	   
	   {\bf Acknowledgement}
	   
	   \vspace{0.3cm} 
	   
	   We acknowledge useful discussions with G. Filatrella, D. Giuliano, P. Lucignano and G. Miano. This work was supported financially by University of Napoli "Federico II" with project PLASMJAC, E62F17000290005.

   \vspace{1.0cm}

\begin{appendix}
   
\section{Fluxon energy in the dissipationless case}
   
  Let us neglect for the time being the electromagnetic source in the motion equation for the fluxon, Eq.(\ref{motion}). The $\gamma$ term accounts for a current source and can sustain the propagation of the fluxon along the junction compensating the dissipations.  In the infinite length limit for the junction an energy eigenmode for the fluxon can be derived. When considering the motion equation, Eq.(\ref{roro3}), the eigenmode will have a dispersion characterized by the $k-$vector $ k_x$. 
The $3-d$ Hamiltonian $H_0$ for the fluxon in the absence of dissipation is:
\bea
H_0 = \frac{\hbar}{2e} J_c \: \lambda _J^2 \int d\tilde x\: \left [ \met \left (  \varphi_{\tilde t } \right )^2 + \met \left (  \varphi_{ \tilde x } \right )^2 + (1 -\cos \varphi )\right ]\nonumber
\enea
with  $ \tilde t  = \omega _J t $ and $ \tilde x = x/ \lambda _J$. We drop the tilde in the following, unless needed.  We drop for the time being the prefactor in front of the integral  which can also be rewritten as  $[ \phi_o^2 / ( 8 \pi ^2 ( 2\lambda _L +d )) ]\times 1/2\pi $.  In the infinite length limit for the junction, let us consider a forward moving fluxon  of  the form of Eq.(\ref{flux}).  The energy of the soliton is easily calculated:
\bea
{\cal{E}}_0^{sol} =  \frac{8}{ \sqrt{ 1-u^2} }, 
\enea
where $u$ is the velocity  in unity of the light velocity $c$,
and is conserved:
\bea
\frac{\partial H_0}{\partial t} = \int dx\: \varphi_t \left [ \varphi_{xx} - \varphi_{tt} -\sin \varphi  \right ] =0\label{der}
\enea

If we neglect the $\beta-$term and we use the form of the dissipationless fluxon of Eq.(\ref{flux}), but with a   perturbed steady state velocity of the fluxon driven by the current $J_{ext}$ ($\gamma = J_{ext}/J_{c}$):
\bea
\alpha \:\frac{ 8 \: u_\infty /{\bar c}}{\sqrt{1-( u_\infty /{\bar c})^2}} = 2\pi \: \gamma,
\:\:\:\:\: 
\left (\frac{u_\infty}{{\bar c}}\right ) ^2 =  \frac{1}{1 + \left (\frac{4\:\alpha }{\pi \gamma} \right )^2},\nonumber\\
 E_0^{sol} = \frac{8}{\sqrt{1-( u_\infty /{\bar c})^2}}= \frac{ 2\pi \gamma {\bar c} }{\alpha \: u_\infty }  = 8 \sqrt{ 1 +  \left (  \frac{ \pi \gamma }{4\: \alpha } \right )^2} 
 \enea
These results are well known\cite{salerno}. 

\vspace{0.5cm}

\section{ Dissipative terms in the absence of excitation modes interaction}
In our approximations we expect that the dissipative terms, in presence of a bias current $\gamma$ lead to a stationary state dynamics which is not qualitatively  different from the dynamics  presented in the dissipationless simulation of Section VI. The dissipation losses should be compensated by the driving current. 
 In the absence of the forcing term, the fluxon velocity $u$ can be determined  following the same lines of  Appendix A, with inclusion of the $\beta-$term.   In analogy with Eq.(\ref{der}),  we impose: 
  \begin{widetext}
 \bea
\alpha  \:\frac{1 }{\omega _J}  \:\int  dx \: ( \partial _t\varphi )^2  - \beta \: \int  dx dx'\:  \left [  \frac{\partial \varphi  }{\partial t } ( x,t) \: G(k(x ,x')) \: 
\frac{\partial^3 \varphi  }{\partial t\partial ^2 x' } ( x',t)\right ]  - \int  dx \:  \gamma \:  \partial _t\varphi =0,
  \label{roro41}
 \enea
 \end{widetext}
 As in the derivation of the SPP, we rely on the fact that the MM  modulation is subwavelength and that all space dependences are on a scale larger that the periodicity $d$ of the groove lattice.  In particular  $k = 2\pi /\lambda_\omega  < <	\pi /d  $, so that we can consider just average homogeneous MM  contacts and the Green's function  $G(k; x, x') $  satisfying Eq.(\ref{rota5}) and vanishing at the junction edge,  takes the simple form 
 \bea
G(k; x, x') = \frac{1}{k\: \sinh k L} \left [ \sinh k ( L-x_>) \:  \sinh k x_ < \right ],
 \enea 
where $ x_> (x_< )$ is the larger ( smaller) argument between $ x,x' $.  Far from the edges, we can approximate the unperturbed kink  $\varphi ( \xi) $ as a step function at $\xi = x-ut = L/2$. Hence,   $ \varphi _t$ has even symmetry in space with respect to $L/2$, while    $ \varphi _{tx}$  has odd symmetry.  On the other hand   $G(k; x ,x')  \sim G(k|x -x'|) $  and a double integration by parts changes the $\beta -$term into
\bea
\int  dx dx'\:  \left [  \frac{\partial \varphi  }{\partial t } ( x,t) \:   \frac{\partial ^2}{\partial x\partial x' } G(k; x ,x')  \: 
\frac{\partial \varphi  }{\partial t } ( x',t)\right ]. 
\enea
Here $  \frac{\partial ^2}{\partial x\partial x' } G(k; x ,x') $ is very localized  at $x\sim x'$, so that this term can be changed into a local term which renormalizes the $\alpha-$term.  The derivation of $u_{\infty} $ given in Appendix A follows. 

\section{Forcing terms in the non dissipative  S-G equation of motion }
   
   Let us now derive the forcing terms to be added in the S-G equation of motion Eq.(\ref{sgor}), assumed to be one-dimensional and  non dissipative ($\alpha ' =0$).  

     From Eq.\ref{cura}, the phase jump between the two edges of the insulating layer is:
     \bea
 \frac{\partial \vartheta _+}{\partial x }  -   \frac{\partial \vartheta _-}{\partial x } &=& \frac{8\pi ^2 \lambda_L^2} {\phi_o c }  \left [ J_+-  J_- \right ] \nonumber\\
 &-& \frac{2e}{\hbar c }\: \left [ A_x( +\infty) -A_x( -\infty) \right ], 
 \label{inoo}
\enea
 $ \left [ J_+-  J_- \right ] $ is the difference of superconducting screening currents at the  barrier boundaries of the Josephson  Junction. Eq.(\ref{inoo})  is consistent with the  London equation Eq.(\ref{lond}b):
      \bea
 \frac{\partial J^{s} }{\partial t }  = \left .  \frac{n_s e^2}{m} \vec{E} \right |_b,    \:\:\:\:\:\:\:\:    \nabla \times \vec{J}^{s} + \frac{n_s e^2}{mc} \vec{B} =0 .
 \label{lond} 
 \enea
 In fact, usually the contacts are bulk superconductors  and the $\vec{A}$ field decays far from the edge on the scale of $\lambda _L$ and it is possible to take a circuit with  $z_\pm$ well within the bulk so that $J_\pm \equiv J(z_\pm ) $ vanish and the circulation of $\vec{A}$ along the circuit provides the full flux piercing  the weak link area  $ ( 2\lambda _L +d_i) \: L \: B_y $.  In the limit to an inhomogeneous but spacially continuous superconductor, $2\lambda _L +d_i \to z_+-z_- \to 0 $,  a local expression can be obtained  and  the finite difference of the currents $J_\mp$ divided by $ z_+-z_- $ turns into   the  $curl$ of the screening currents: 
      \bea
     \frac{ [ J_- - J_+ ]}{ 2\lambda _L +d_i} \to  - \partial_z J^{sb}_x \sim  - \left . \nabla \times \vec{J}^{sb} \right |_y\nonumber\\
  \label{trick}
  \enea
(the label $b$ stands for "bulk").   Similarly, for a continuous phase 
 \bea
    - \frac{  \frac{\partial \vartheta _+}{\partial x }  -   \frac{\partial \vartheta _-}{\partial x }  }{ 2\lambda _L +d_i}   \to  \left . \vec{\nabla} \times \vec{\nabla}\vartheta\right |_y =0.\nonumber\\
  \label{trick}
  \enea
  so that,  the $\hat y$  component of the London equation, Eq.(\ref{cura}), is recovered ( $1/ \lambda _L^2 =   (4\pi n_s e^2/ mc)$):
  \bea
  0= -  \left . \nabla \times \vec{J}^{sb} \right |_y  - (n_s e^2/ mc)\: \left .  B_y \right |_b.
  \label{lolo}
  \enea
  However, there are two crucial differences in our case. On the one hand the thickness of the superconducting contacts in the overlap junction is finite and relatively small. On the other hand, there is the SPP leaking into the upper superconducting edge generated by the  MM at the top,  which  does not allow to drop the difference in the current flowing between the two contacts.  We assume that the perturbed phase difference $\varphi $ depends on the current imbalance induced by the plasmon and on the source  field, $B^{source}_y$, which penetrates a distance $\lambda _p$ along the $\hat z$ direction ($\lambda _p$  is discussed in the text). It follows that Eq.(\ref{inoo}), along the $\hat x $ direction, takes the form:
 \begin{widetext}
\bea 
\frac{\phi _0}{2\pi }\:   \frac{\partial \varphi }{\partial x } = - ({ 2\lambda _L +d_i})\: B_y + \frac{4\pi  \lambda_L^2} {c }  \left [ J_+-  J_- \right ] _{SPP}-  \lambda _p \: B_y ^{source}, 
    \label{ino2}
    \end{eqnarray} 
    \end{widetext}
  where we assume that $B_y = B ^{(1)} +B ^{(2)} $, where $B ^{(1)}$ is  the one generated by the fluxon in the absence of the external source and $B ^{(2)}$ is  the one giving rise to radiating effects. 
  
    In the following we derive an expression for  $ curl\: B ^{(2)}$,  which contributes  to the motion equation of  Eq.(\ref{sgor}) with a dissipative term, the usually  called ' $\beta $ term ' (see Eq.(\ref{difg})).
 We will  drop the magnetic field generated by the Josephson current itself in the derivation, which is usually considered to be small.  We have, excluding the term  due to the source,  $B^{source}_y$, for the time being:
 \begin{widetext}
   \bea
\left . \nabla \times \nabla \times \vec{B^{(2)}} \right |_y = \partial_z \left ( \left .  \nabla \times \vec{B^{(2)}} \right |_x\right ) -\partial_x \left ( \left .  \nabla \times \vec{B^{(2)}} \right |_z\right )  = \frac{\partial}{\partial z }\left [  \frac{4\pi }{c}  J^{b}_x + \frac{4\pi \lambda_L ^2 }{c}   \left (   \frac{4\pi \: i \: \sigma _b\: \omega }{c^2} + \epsilon_\infty \frac{\omega ^2}{c^2} \right ) {J} ^{b}_x 
 \right ] \nonumber\\
 - \frac{\partial}{\partial x }\left [ \frac{4\pi}{c}  \left . J_z\right |_T\right ]
 \label{Ctota}
\enea
\end{widetext}
Here $ {J} ^{b}_x $ denotes the superconducting screening currents induced by the radiation at the boundary of the contacts.  The first square bracket term on the r.h.s. arises from the Maxwell-Ampere equation
\bea
 \left .  \nabla \times \vec{B^{(2)}} \right |_x =  \frac{4\pi }{c}  J^{}_x + \frac{1}{c} \frac{d  {D}^{(b)}_x}{dt},
 \label{Cma}
 \enea
 where  the total current $J$ also includes a contribution from  the Ohmic transport, $ J= J^{b}  + J^{nb}  =  J^{b}+ \sigma _b  \: E _b^{(2)}$.  $ \vec{D}^{(b)} $ is the electric induction vector penetrating  in the contacts. Both $ E _{b x}^{(2)} $ and $ {D}_x^{(b)} $ of  Eq.(\ref{Cma}) can be related to $J_x^{b}$ itself, by means of the Fourier transform ($ \partial _t \to -i\:\omega $)  of the  London equation, Eq.(\ref{londi}):  $E _{b x}^{(2)}= - {4\pi \lambda_L ^2 \: i \: \omega\: J_x ^{b} }/{c^2} $.
 
  The second square bracket term   on the r.h.s. of Eq.(\ref{Ctota})  accounts for  the normal quasiparticle  tunnelling current, $ \left .   \vec{J} _n  \right |_{T}$,  oriented along $\hat z$.  Quasiparticles  excited due to the high  operating frequency ($ \omega >> \Delta /\hbar$) contribute dissipatively to the current.  The quasiparticle current coexists with the Josephson current  $J_c \sin \varphi $. $\sigma _{qp}$ is the corresponding quasiparticle conductivity and  this term can be expressed in terms of the derivatives of the phase difference $\varphi$: 
 
 \begin{widetext}
\bea
    \frac{4\pi}{c} \frac{\partial J_z^{nT} }{\partial x }  =  \frac{4\pi}{c} \sigma _{qp}\frac{\partial E_z  }{\partial x } \approx  \frac{4\pi}{c} \frac{\sigma _{qp}}{d}\frac{\partial V(z=0) }{\partial x } = \frac{4\pi }{c}  \:\sigma _{qp}\: \frac{\hbar}{2e \: d}\frac{\partial^2 \varphi  }{\partial t\partial x }. \nonumber
       \enea 
\end{widetext}
Finally,  according to Eq.(\ref{lolo}), 
  \bea
 \left (  \frac{4\pi}{c} \nabla \times  J^{sb} \right )_y  \approx    \frac{4\pi}{c} \partial _z J^{sb} _x   =   - \frac{1}{ \lambda _L^2}  \: B_y ^{(2)}, 
\enea

so that  Eq.(\ref{Ctota}) can be written as\cite{bulaevskii}:
\bea
\left . \nabla \times \nabla \times \vec{B}^{(2)} \right |_y + \frac{1}{ \lambda_\omega  ^2}  \: B_y^{(2)} =-  \frac{4\pi}{c} \frac{ \hbar }{2e\: d} \: \sigma _{qp}\frac{\partial^2 \varphi  }{\partial t\partial x }, 
\label{Crota6}\\
{\rm with}\:\:\:  \frac{ 1}{ \lambda_\omega  ^2}   \equiv  \frac{1}{ \lambda_L ^2} +  \epsilon _{b}(\omega )  \frac{\omega ^2}{c^2} \hspace{3cm}
     \nonumber\\
  {\rm and }\:\:\:\:\:\:  \epsilon (\omega )  = \epsilon_\infty +   \frac{4\pi \: i \: \sigma _{b} }{c^2\: \omega}. \hspace{3cm} \nonumber
 \enea
The  full motion equation for the fluxon  is reported in the text,  Eq.(\ref{roro3}).

\end{appendix}
 
  \end{document}